\documentclass[fleqn,usenatbib]{mnras}

\usepackage{newtxtext,newtxmath}

\usepackage[T1]{fontenc}
\usepackage{ae,aecompl}

\usepackage{graphicx,epstopdf}	
\usepackage{amsmath}	
\usepackage{amssymb}	
\usepackage{upgreek}

\defcitealias{GinzburgSari2018}{GS18}

\title[The End of Runaway]{The End of Runaway: How Gap Opening Limits the\\ Final Masses of Gas Giants}

\author[S. Ginzburg and E. Chiang]{
Sivan Ginzburg$^{1,2}$\thanks{E-mail: ginzburg@berkeley.edu}
and Eugene Chiang$^{1,3}$
\\
$^{1}$Department of Astronomy, University of California at Berkeley, CA 94720, USA\\
$^{3}$Department of Earth and Planetary Science, University of California at Berkeley, CA 94720, USA
}

\date{Accepted XXX. Received YYY; in original form ZZZ}

\pubyear{2019}

\begin{document}
\label{firstpage}
\pagerange{\pageref{firstpage}--\pageref{lastpage}}
\maketitle

\begin{abstract}
Gas giants are thought to form
by runaway accretion: 
an instability driven by the self-gravity of growing
atmospheres that causes accretion rates to rise
super-linearly with planet mass.
Why runaway should stop at a Jupiter or
any other mass is unknown. We consider the proposal
that final masses are controlled by 
circumstellar disc gaps (cavities) 
opened by planetary gravitational torques.
We develop a fully time-dependent theory of gap formation
and couple it self-consistently to planetary growth rates.
When gaps first open, planetary torques overwhelm viscous torques,
and gas depletes as if it were inviscid.
In low-viscosity discs, of the kind motivated by recent observations
and theory, gaps stay predominantly in this inviscid phase
and planet masses finalize at $M_{\rm final}/M_\star\sim(\Omega t_{\rm disc})^{0.07}(H/a)^{2.73}(G\rho_0/\Omega^2)^{1/3}$, with 
$M_\star$ the host stellar mass,
$\Omega$ the planet's orbital angular velocity,
$t_{\rm disc}$ the gas disc's lifetime,
$H/a$ its aspect ratio,
and $\rho_0$ its unperturbed density.
This final mass 
is independent of 
the dimensionless viscosity 
$\alpha$ and applies to large orbital
distances, typically beyond $\sim$10 AU, 
where disc scale heights exceed planet radii.
It evaluates to 
a few Jupiter masses at 10--100 AU, increasing
gradually with distance as gaps become harder to open.
\end{abstract}




\section{Introduction}\label{sec:introduction}

\footnotetext[2]{51 Pegasi b Fellow}

Perhaps
the leading theory
for the formation of giant planets is the core accretion model \citep{PerriCameron1974,Harris1978,Mizuno1978,Mizuno1980,Stevenson1982,BodenheimerPollack86,Pollack96}. According to this model, 
gas giants assemble from the bottom up:
a rocky or icy core several times the mass of the Earth
accretes gas from the circumstellar disc in which it is embedded.
The process must complete within a few million years,
the observed disc lifetime 
\citep{Mamajek2009,WilliamsCieza2011,Alexander2014}.
Initially, gas accretion is regulated by Kelvin--Helmholtz cooling, i.e., how long it takes the nascent atmosphere to radiate away its gravitational
potential
energy \citep{PisoYoudin2014}.
When still lighter than the core,
the atmosphere 
grows (read: doubles in mass)
on a timescale that increases as roughly the square of
its mass \citep{LeeChiang2015}.
The situation changes 
dramatically 
once the atmosphere
exceeds
the core in mass.
Then the self-gravity of the atmosphere
causes it to cool---and therefore accrete mass from the surrounding disc---over ever shorter timescales. Gas accretion
is now in the ``runaway'' phase
\citep[][and references therein]{BodenheimerPollack86,Pollack96,Ikoma2000,Lee2014,PisoYoudin2014,LeeChiang2015,Piso2015,Berardo2017}.

How does runaway accretion stop?
In other words, what determines the final mass
of a gas giant? The total gas disc mass sets a strict 
upper bound, but for most discs this is not
constraining, as inside an orbital radius
of a few hundred AU, discs are estimated to 
have 10--100 times more gas than Jupiter 
\cite[see, e.g., Fig. 10 of][]{Tripathi2017}.
Even these gas mass estimates are often only lower bounds
because they are based on masses inferred from
dust emission and gas-to-dust ratios
that are assumed to be solar; in reality,
inward radial drift of solid particles leads to
higher gas-to-dust abundances at large disc radii 
\citep[e.g.,][]{Andrews2012,Powell2017}. 
The problem of stopping runaway is made more acute by
recent indications that 
the giant planet mass function is
bottom-heavy, with the number of planets per decade
in mass decreasing inversely with mass 
\citep{Nielsen2019,Wagner2019}. Understanding the endgame of giant planet
formation---in particular the accretion luminosities
of nascent giants \citep[e.g.,][]{Berardo2017}---is critical
for motivating and interpreting observational campaigns
to directly image protoplanets.

Gap opening by planets \citep{GoldreichTremaine80}
has been considered a key 
mechanism
halting a
planet's growth
\citep{LinPapaloizou93,Bryden99,Kley99,Lubow99,DAngelo03,TanigawaIkoma2007,Lissauer2009,Machida2010,TanigawaTanaka16}.
Gravitational torques exerted by the planet
repel gas away from its orbit,
carving a gap (annular cavity) in the disc and
diminishing the supply of gas
to the planet. At the same time,
viscous torques intrinsic to the disc diffuse material
back into the gap. The essential quantity to
compute is
the gap depth, i.e., the extent to which the gas density
in the planet's vicinity is lowered relative to the background 
disc value \citep{DuffellMacFadyen13,Fung2014,Kanagawa2015}.\footnote{\citeauthor{Kratter2010} (\citeyear{Kratter2010}; see also \citealt{Thorngren2016} and \citealt{RosenthalMurrayClay2018})
suggested in their subsection 5.2.1 an ad-hoc gap starvation mass
based on gap widths. Their discussion motivates a more
thorough analysis of accretion rates inside gaps, which is what
our study provides.} 
\citet[][hereafter \citetalias{GinzburgSari2018}]{GinzburgSari2018} calculate both gap depths 
and radial widths by considering in detail how planet-driven waves
dissipate within gaps. In addition
to providing physical justifications for previous numerical results,
they uncover new gap scaling behaviours appropriate to low
disc viscosities.

Here we build upon these 
advancements
to re-visit 
the role of gap opening
in determining the final masses of gas giants.
We focus on how gaps open in discs having low viscosities
($\alpha \leq 10^{-3}$,
where $\alpha$ is the turbulent Mach number introduced
by \citealt{ShakuraSunyaev73}), as motivated by 
Atacama Large Millimeter Array (ALMA) observations 
that not only reveal gaps in discs but also point to nearly inviscid environments 
\citep[e.g.,][]{Pinte2016,Flaherty2017,Dong2018,HartmannBae2018,Zhang2018}. 
Another novel feature of our analysis is that
we account for how gap depths vary (deepen) with time.
Our approach is analytic and approximate, to gain
intuition and guide future, more precise numerical studies.

The outline of the rest of this paper is as follows.
After setting down
in the next subsections 
our model assumptions,
we describe in Section \ref{sec:gapless}
how runaway growth proceeds unchecked when gap opening is ignored.
The tables turn 
in Section \ref{sec:gap} where
we account for gap clearing and provide closed-form expressions
for the final masses of gas giants. In Section \ref{sec:conclusions}
we summarize and 
give an outlook.

\subsection{Accretion onto sub-thermal planets:\\Spherical symmetry at the Bondi radius}\label{sec:spherical}

We restrict our analysis to planets
with masses $M$ below the thermal mass, 
defined here as that for which the
Hill radius 
$R_{\rm H}$
equals the gas disc's scale height $H$:
\begin{equation} \label{eq:mass_thermal}
M_{\rm thermal} \equiv 3 (H/a)^3 M_\star 
\end{equation}
where $M_\star$ is the host star mass and $a$ is the orbital radius.
Historically, only planets above $M_{\rm thermal}$ were thought 
to open gaps 
and stop growing \citep[e.g.,][]{LinPapaloizou93,Bryden99}, the rationale being that
planet-driven density waves 
had to be
non-linear in order to shock, dissipate, and transfer angular momentum to the disc. However, \citet{GoodmanRafikov2001} explained that even
linear 
waves gradually steepen and eventually shock as they propagate away from the  planet.
Modern numerical calculations \citep{DuffellMacFadyen13,FungChiang17}
confirm 
that sub-thermal planets are indeed capable of carving out deep gaps (if $\alpha$ is sufficiently low) and
can therefore
potentially stop growing before reaching $M_{\rm thermal}$.

Sub-thermal planets are easier to model than super-thermal planets
in the following sense.
In general, the outer radius of a planet embedded in a disc
is the smaller of the Hill radius or the Bondi radius
\begin{equation}
R_{\rm B} = GM/c_{\rm s}^2 
\end{equation}
where $c_{\rm s}$ is the nebular sound speed and $G$ is the gravitational
constant. 
For
$M < M_{\rm thermal}$,
the length scales order as
$R_{\rm B}<R_{\rm H}<H$.
This hierarchy simplifies planetary accretion in several ways. First, because
sub-thermal planetary atmospheres 
are on scale $R_{\rm B}< H$, anisotropies in disc density (differences in
density between the vertical and in-plane directions)
can be ignored; the atmospheres will have a roughly
spherical symmetry
\citep{Rafikov2006,PisoYoudin2014}.
Second, most of a planet's repulsive torque is carried by waves launched at radial distances $\sim$$H$
away from the planet \citep{GoldreichTremaine80}
and deposited at
still larger distances \citep{GoodmanRafikov2001}.
Thus, the gas nearest the planet
(at the bottom of the
gap, which even for a sub-thermal planet can be deep) 
has an approximately flat radial density profile,
up to distances of at least $H>R_{\rm B}$ \citep[e.g.,][]{Kanagawa2015}. 
Finally, the fact that $R_{\rm B}<R_{\rm H}$ means that the Keplerian shear velocity at the planet's outer Bondi radius is subsonic
and smaller than the circumplanetary orbital velocity,
supporting the classical Bondi picture which neglects
angular momentum and takes the accretion velocity
to be sonic at $R_{\rm B}$.
In other words, a circumplanetary disc does not necessarily form at $R_{\rm B}$, and accretion on that
scale may
be spherically symmetric (rotation becomes
significant on scales
$< R_{\rm B}$; see our Section \ref{sec:caveats}).

Our restriction to sub-thermal masses and assumption of spherical symmetry 
mean that we are working in a regime
complementary to that considered by \citet{Szulagyi2014},
who address the problem of limiting giant planet masses
in the context of circumplanetary discs \citep[see also][]{SzulagyiMordasini2017}.

\subsection{Background disc model}\label{sec:disc}

For most of our calculations we focus on the nominal case of a
nascent giant planet
located at an orbital separation of $a=10\textrm{ AU}$ from a solar mass
star ($M_\star=M_\odot$).
The gas surface density of our disc when unperturbed (i.e., with no 
planet) resembles that of the minimum-mass solar nebula (MMSN)\footnote{
By construction, the MMSN contains just enough gas to form a Jupiter-mass planet. This total mass constraint 
does not appear in our calculations, as we wish to explore other limitations to runaway growth.} 
\citep{Weidenschilling77, Hayashi1981, ChiangYoudin2010}: 
\begin{equation}
\Sigma_0=70\textrm{ g cm}^{-2}\left(\frac{a}{10\textrm{ AU}}\right)^{-3/2} \,.
\end{equation}
The disc temperature 
is similar to that derived by
\cite{ChiangGoldreich97}:
\begin{equation}
T_0=100\textrm{ K}\left(\frac{a}{10\textrm{ AU}}\right)^{-3/7} \,.   
\end{equation}
The sound speed is given by $c_{\rm s}=\sqrt{\gamma k_{\rm B}T_0/\mu}$, where $k_{\rm B}$ is the Boltzmann constant. We take an adiabatic index $\gamma=7/5$ and a molecular weight $\mu=2\textrm{ amu}$
appropriate
for molecular hydrogen. 
The resulting volumetric gas density at the disc midplane
is 
\begin{equation}\label{eq:disc_rho}
\rho_0=\frac{1}{\sqrt{2\uppi}}\frac{\Sigma_0}{H}
=
3\times 10^{-12}\textrm{ g cm}^{-3}\left(\frac{a}{10\textrm{ AU}}\right)^{-39/14}   
\end{equation}
\citep[e.g.,][]{FKR2002} 
with $H=\Omega^{-1}\sqrt{k_{\rm B}T_0/\mu}$ 
the disc's scale height and $\Omega=(GM_{\star}/a^3)^{1/2}$ the orbital angular velocity. For our parameter 
choices, the disc's aspect ratio equals
\begin{equation}
H/a = 0.08 \left( \frac{a}{10 \textrm{ AU}} \right)^{2/7} \,.
\end{equation}
Our nominal disc is thicker (hotter) than in 
other models (e.g., \citealt{DAlessio1998} compute $H/a\approx 0.06$ in their Fig. 6), possibly
causing us to overestimate, by factors of a few, 
both the range of planet masses
that can remain sub-thermal (equation \ref{eq:mass_thermal}) 
and final planet masses (Section \ref{sec:final},
in particular equation \ref{eq:final_mass}).
We allow ourselves this latitude as there are
other model uncertainties that are also order-unity
if not larger.

We assume a nominal disc lifetime of $t_{\rm disc}=3\textrm{ Myr}$, during which the background density $\Sigma_0$ remains constant, and after which it vanishes.
This time covers only the runaway phase of
planetary gas accretion and its aftermath; the assembly of the
underlying rocky core, and the initial pre-runaway
phase of gas accretion (discussed briefly in
Section \ref{sec:cooling}) 
are not covered. What fraction of time rocky core assembly
takes of the total disc
lifetime (spanning 1--10 Myr; e.g., \citealt{Mamajek2009}; 
\citealt{Pfalzner2014}) is unknown; as such,
we may be overestimating the duration
of the runaway/post-runaway phase
when setting it equal to $t_{\rm disc} = 3\textrm{ Myr}$.
Note that merely decreasing $t_{\rm disc}$ in order
to stop runaway at some desired mass would require unreasonable
fine-tuning, as growth timescales become extremely
short with increasing mass 
(it is, after all, called ``runaway'' for a reason;
see, e.g., equation \ref{eq:t_cool}).
We will see in our full theory of accretion within gaps
that final planet masses
are insensitive to changes in $t_{\rm disc}$,
changing only by several percent
when $t_{\rm disc}$ varies by factors of a few
(equation \ref{eq:final_mass}).

\section{Gapless accretion}\label{sec:gapless}
We first consider how planets accrete gas from discs neglecting
gap clearing.
This will demonstrate the problem of runaway accretion in both thermodynamic (Section \ref{sec:cooling}) and
hydrodynamic (Section \ref{sec:Bondi}) senses.

\subsection{Runaway cooling}\label{sec:cooling}

Pre-runaway gas accretion onto planetary cores has been studied extensively, both in the context of gas giants and short-period sub-Neptunes \citep{Pollack96,Ikoma2000,PapaloizouNelson2005,Rafikov2006,PisoYoudin2014,Lee2014,LeeChiang2015,Piso2015,Ginzburg2016}. These studies find that the planet's atmosphere is divided into an interior convective region 
containing most of the mass,
and an outer radiative and nearly isothermal envelope which extends to $R_{\rm B}$.
Conditions at the radiative--convective boundary dictate the planet's cooling rate which is the bottleneck for pre-runaway accretion. The atmosphere grows (doubles) on a timescale that increases with its mass as long as it is much lighter than the core. Once the atmosphere and core become comparable in mass, the growth time decreases, initiating the runaway phase.

In this subsection we extend the analytical scaling relations of previous studies to the runaway phase, where the core mass can be neglected. The main difference is that the gravity field is dominated by the gas itself, and not by the core, transforming the atmosphere's density profile in the convective region into a polytrope
akin to those in stars.
We keep the 
following analysis
approximate,
as we shall see that the cooling rate derived in this subsection is not the
ultimate bottleneck for accretion.

The density in the convective region scales with
temperature as $\rho(T)/\rho_{\rm rcb}=(T/T_0)^{1/(\gamma-1)}$, with $\rho_{\rm rcb}$ denoting the density at the radiative--convective boundary (RCB), and $T_0\sim T_{\rm rcb}$ since the radiative envelope is isothermal. From hydrostatic equilibrium, the temperature at the planet's centre is given by $k_{\rm B}T_{\rm c}\sim GM\mu/R_{\rm rcb}$, with $M$ denoting the planet's mass (assumed dominated by the convective gas layer), and $R_{\rm rcb}$ the radius of the convective polytrope. By substituting $\rho(T_{\rm c})\sim M/R_{\rm rcb}^3$ for the central density of the polytrope we find
\begin{equation}\label{eq:rho_rcb}
\rho_{\rm rcb}=\frac{M}{R_{\rm rcb}^3}\left(\frac{k_{\rm B}T_0R_{\rm rcb}}{GM\mu}\right)^{1/(\gamma-1)}=\frac{M}{R_{\rm rcb}^3}\left(\frac{R_{\rm rcb}}{R_{\rm B}}\right)^{1/(\gamma-1)}.
\end{equation}
The luminosity is calculated by applying the diffusion equation to the radiative layer
$L\sim\sigma T_0^4R_{\rm B}/(\kappa\rho_{\rm rcb})$, where $\sigma$ is the Stefan--Boltzmann constant and $\kappa$ is the opacity at the RCB
(see, e.g., Section 2.2 of \citealt{Ginzburg2016}; we have here omitted
order-unity coefficients).
The gravitational energy that the planet has to radiate away is 
$E\sim GM^2/R_{\rm rcb}$. The Kelvin--Helmholtz cooling time is therefore
\begin{equation}\label{eq:t_cool_rcb}
t_{\rm cool}\sim\frac{E}{L}\sim\left(\frac{k_{\rm B}}{\mu}\right)^5\frac{\kappa}{\sigma G^4}\frac{T_0}{M^2}\left(\frac{R_{\rm B}}{R_{\rm rcb}}\right)^{(4\gamma-5)/(\gamma-1)},
\end{equation} 
where we substituted $\rho_{\rm rcb}$ from equation \eqref{eq:rho_rcb}.
The radius
$R_{\rm rcb}$ is related to $R_{\rm B}$ by a logarithmic factor that stems from the exponentially declining density profile in the outer isothermal envelope \citep{PisoYoudin2014,Ginzburg2016}. 
We omit this logarithmic factor and approximate $R_{\rm rcb}\sim R_{\rm B}$, simplifying equation \eqref{eq:t_cool_rcb}:
\begin{equation}\label{eq:t_cool}
t_{\rm cool}
\sim
\left(\frac{k_{\rm B}}{\mu}\right)^5\frac{\kappa}{\sigma G^4}\frac{T_0}{M^2}
\approx
10^4\textrm{ yr } \left(\frac{T_0}{100\textrm{ K}}\right)\left(\frac{M}{M_{\rm J}}\right)^{-2},
\end{equation}
where $M_{\rm J}$ is the mass of Jupiter and we have assumed for simplicity a constant $\kappa=0.1\textrm{ cm}^2\textrm{ g}^{-1}$ 
(from dust; see Appendix C in \citealt{Piso2015}).

Equation \eqref{eq:t_cool} demonstrates how decreasing cooling times lead to runaway growth on a timescale much shorter than the disc's lifetime. One interesting feature of runaway cooling is its 
near-insensitivity to the nebula's density.
Pre-runaway cooling is characterized by a similar
insensitivity \citep{LeeChiang2016}.

We emphasize that equation \eqref{eq:t_cool}, which indicates that
the cooling/growth timescale decreases with increasing envelope mass,
is valid only for atmospheres that outweigh their cores, $M>M_{\rm core}$.
By contrast, lighter atmospheres are accreted on 
timescales that increase with increasing gas mass \citep{LeeChiang2015}.
We conclude that atmospheres take the longest time to double their mass when $M\sim M_{\rm core}$. By setting the cooling/growth time in equation \eqref{eq:t_cool} to the gas disc lifetime of $\sim$$3\times 10^6$ yr, 
we see that only 
planets above a critical mass of $M \sim
10$--$20\,M_\oplus$ 
achieve runaway, for our nominal 
$T_0=100\textrm{ K}$ (see \citealt{PisoYoudin2014} and \citealt{Piso2015} for a more comprehensive analysis). Our derivation is indicated graphically in Fig.  
\ref{fig:growth}. The intersection (just outside and to the left of the plotted range) of the cooling timescale
(dashed green line) and the gas disc lifetime (dotted horizontal black line) provides a rough estimate for the critical 
planet mass (and by extension core mass $M_{\rm core}\sim M$).

\subsection{Bondi accretion}\label{sec:Bondi}

As is apparent from equation \eqref{eq:t_cool} and Fig.~\ref{fig:growth}, the cooling timescale becomes exceedingly short during runaway. At some stage, the nebula will not be able to replenish gas at a sufficient rate to the planet's outer boundary ($R_{\rm B}$) to maintain this rapid growth. Accretion ceases to be cooling/thermodynamically limited and becomes instead hydrodynamically limited.

The maximum rate at which gas can be brought to the Bondi radius is given by the free-fall velocity, which is also the sound speed at that radius. The corresponding mass accretion rate is the Bondi rate, $\dot{M}=4\uppi R_{\rm B}^2c_{\rm s}\rho$, where $\rho$ is the ambient gas density. The fastest, Bondi-limited growth time is then
\begin{equation}\label{eq:t_bondi}
t_{\rm Bondi}\equiv\frac{M}{\dot{M}}=\frac{c_{\rm s}^3}{4\uppi G^2\rho M},
\end{equation} 
which, like $t_{\rm cool}$, decreases with increasing $M$ and therefore also implies runaway behaviour.

In an attempt to calculate the final masses of gas giants, \citet{TanigawaTanaka16} utilize an accretion formula
that is drawn empirically 
from two-dimensional simulations by \citet{TanigawaWatanabe2002} 
and scales as $\dot{M} \propto M^{4/3}$ \citep[recently this scaling has been explained analytically by][in terms of isothermal accretion shocks]{Lee2019}.
These simulations account for hydrodynamical flows 
that become increasingly complex and rotation-dominated 
as the planet mass increases.
However, in the limit $R_{\rm B}\ll R_{\rm H}$, i.e., $M\ll M_{\rm thermal}$, the geometry and flow are expected to be simple
and to match the results of classical Bondi accretion:
the envelope boundary at $R_{\rm B}<H$ should be more-or-less
spherically symmetric, and the Keplerian shear velocity
at $R_{\rm B}$ should
be negligible as it is smaller than the sound speed $c_{\rm s}$ (see our Section \ref{sec:spherical}).
Indeed, as can be appreciated from Fig.~1 of \citet{TanigawaTanaka16},
the numerically simulated three-dimensional accretion rates at the lowest masses
match the Bondi scaling $\dot{M}\propto M^2$ better than 
$\dot{M}\propto M^{4/3}$.
As we assume in this paper that $M < M_{\rm thermal}$---an
assumption that becomes better justified at
larger orbital distances---we use the Bondi result (see also the discussion in Sections \ref{sec:spherical} and \ref{sec:caveats}).

If we ignore gap clearing, the surrounding density is given by the unperturbed nebular value $\rho=\rho_0\approx 3\times 10^{-12}\textrm{ g cm}^{-3}$ at our nominal disc radius
of 10 AU (Section \ref{sec:disc}). By comparing equations \eqref{eq:t_cool} and \eqref{eq:t_bondi} using this background (gapless) $\rho$, we find that the Bondi rate limits gas accretion only at unrealistically high masses---above $200\, M_{\rm J}$ for our nominal 100 K disc (and even then, accretion still proceeds in a runaway manner, as explained above). In the next section, we  consider how gaps lower $\rho$ and stop 
runaway at smaller masses.

\section{Gap Clearing}\label{sec:gap}

Equation \eqref{eq:t_bondi} indicates that the Bondi-limited growth time depends on the planet's ambient density $\rho$ (in contrast to the Kelvin--Helmholtz cooling time, which is insensitive to $\rho$;
see Section \ref{sec:cooling}). 
Therefore, once a planet opens a gap around its orbit,
the decreasing density prolongs the Bondi accretion timescale,
conceivably beyond the gas disc lifetime $t_{\rm disc}$.

By balancing gravitational 
and viscous torques, 
\cite{Fung2014} derived an analytical scaling for the depth of a gap with respect to the background: $\rho/\rho_0\sim\alpha h^5/m^2$, where $h\equiv H/a$ is the disc's aspect ratio and $m\equiv M/M_\star$ is the planet-to-star mass ratio (see also \citealt{DuffellMacFadyen13} who found the same scaling empirically). 
The \citet{ShakuraSunyaev73} $\alpha$ parametrizes the kinematic viscosity $\nu\equiv\alpha c_{\rm s}H$. The scaling compares well against
some
multidimensional hydrodynamical simulations
\citep[e.g.,][]{Kanagawa2015n,FungChiang16}
and demonstrates that even low-mass planets can open deep gaps, if the viscosity is low enough. By substituting
the gap density scaling
into equation \eqref{eq:t_bondi}, we find that $t_{\rm Bondi}\propto m$, implying that gap clearing 
stops accretion from running away. 
Eventually, the planet will grow to its final mass, for which $t_{\rm Bondi}$ becomes as long as 
$t_{\rm disc}$.
\citet{TanigawaTanaka16} also\footnote{We reiterate that \citet{TanigawaTanaka16} use an empirical accretion rate
$\dot{M} \propto M^{4/3}$, whereas we adopt the 
Bondi rate $\dot{M} \propto M^2$.
Our choice is justified in Section \ref{sec:Bondi}.} used this analytical depth scaling to argue that gap opening can limit the final mass of gas giants to about $10\, M_{\rm J}$, a result that depends on the viscosity parameter $\alpha$.
  
A new analysis by \citetalias{GinzburgSari2018} indicates that the 
above scaling for $\rho/\rho_0$
is altered for very low viscosities. The reason is that the total gravitational torque that the planet exerts
is no longer dominated by interaction with gas at a distance $H$ from the planet \cite[as assumed in the derivation of][]{Fung2014}. Once the density there drops sufficiently, the total torque becomes dominated by interaction with gas farther away from the planet, leading to an even 
wider and
deeper gap. 
At the same time, 
\citetalias{GinzburgSari2018} also find that the time to open a gap in a low-viscosity disc is long---so long that 
planets might not have enough time to fully clear their gaps as they grow.

Here we calculate the final mass of a gap-opening planet by taking into account the updated gap profiles from \citetalias{GinzburgSari2018} and the temporal evolution of the gap clearing process. This new analysis leads to qualitatively different results. 

\subsection{Simplified solution}\label{sec:intuition}

We first develop intuition by deriving in a simplistic way the final mass of a gap-opening planet in the inviscid ($\alpha=0$) limit.
Our aim here is purely pedagogical; the simplified derivation is inaccurate but captures some of the essential arguments. We present it 
merely as a guide for our more careful
analysis in Section \ref{sec:full}.

We adopt the \citetalias{GinzburgSari2018} dimensionless notation 
$G=M_\star=a=1$. In these units time $t$ is measured in units of
the inverse Kepler frequency $\Omega^{-1} = 1$. 
As before, $m \equiv M/M_\star$ and $h \equiv H/a$.
We rewrite the (now dimensionless) 
Bondi accretion timescale given by equation \eqref{eq:t_bondi} as
\begin{equation}\label{eq:bondi_t0}
t_{\rm Bondi} =
\frac{m}{\dot{m}}
=\frac{\Omega c_{\rm s}^3}{4\uppi G^2\rho_0 M_\star}\frac{\rho_0}{\rho}m^{-1}
= t_0h^3\frac{\rho_0}{\rho}m^{-1} 
\end{equation}
where the dimensionless time
\begin{equation}\label{eq:t0}
t_0\equiv\frac{\Omega^2}{4\uppi G\rho_0}
\end{equation}
is a function of the background density $\rho_0$ and the distance from the star (through 
$\rho_0$ and $\Omega$; $t_0$ is related to the unperturbed nebula's Toomre stability parameter $Q\gtrsim 1$ by an order-unity coefficient).

We now consider how $\rho_0/\rho$, i.e., the depth of the gap
cleared by the planet, evolves with time.
The planet exerts a repulsive torque
$\Sigma m^2/x^3$,
where $x$ denotes the radial distance away from the planet where
the strongest resonant interactions with the disc occur,
inside the gap 
having 
characteristic surface density $\Sigma$
\citep{GoldreichTremaine80,GoodmanRafikov2001,Fung2014,GinzburgSari2018}. 
The torque displaces gas by means of density waves that travel 
a radial distance $w \gtrsim x$ before damping and releasing the angular momentum they carry to the disc \citep{GoodmanRafikov2001,GinzburgSari2018}.
Just outside of $w$, the surface density returns to its unperturbed\footnote{\citetalias{GinzburgSari2018} use a different notation, with $\Sigma_0$ denoting the density at the bottom of the gap and $\Sigma_\infty$ the unperturbed density at its top.} value $\Sigma_0$. 
The angular momentum required to displace an annulus
of width $w \ll 1$ over its own width is $\Sigma_0 w^2$.
In the absence of viscosity, 
the depth and width of the gap grow freely with time as the
planet deposits ever more angular momentum to the disc: 
\begin{equation}
\frac{\Sigma m^2}{x^3}t=\Sigma_0 w^2 \,.
\end{equation}
In later sections and the Appendix, we
consider the physics underlying $x$ and $w$, in particular
how both increase with time, 
and the relation between $w$ and $\Sigma$. 
These time dependencies, which
are necessary for a self-consistent analysis,
are ignored here for simplicity; we assume for now that 
$x \sim w \sim h$, following
previous crude estimates
\citep[e.g.,][]{DuffellMacFadyen13,Fung2014}.
Then the gap contrast increases as
\begin{equation}\label{eq:rho_simple}
\frac{\rho_0}{\rho}=\frac{\Sigma_0}{\Sigma}=\frac{m^2}{x^3w^2}t\approx\frac{m^2}{h^5}t
\end{equation}
from a minimum value of $\rho_0/\rho=1$.

Inserting equation \eqref{eq:rho_simple} into equation \eqref{eq:bondi_t0}, we find
that the planet's mass grows logarithmically with time
(since $\dot{m}\propto\rho\propto t^{-1}$) and reaches a final mass of 
\begin{equation}\label{eq:simple}
m_{\rm final}^{\rm simple}\equiv\frac{M_{\rm final}^{\rm simple}}{M_\star}\sim \frac{h^2}{t_0}=\left(\frac{H}{a}\right)^2\frac{4\uppi G\rho_0}{\Omega^2}
\end{equation}
modulo a logarithmic factor that depends on the gas disc's lifetime
$t_{\rm disc}$.

\begin{figure}
	\includegraphics[width=\columnwidth]{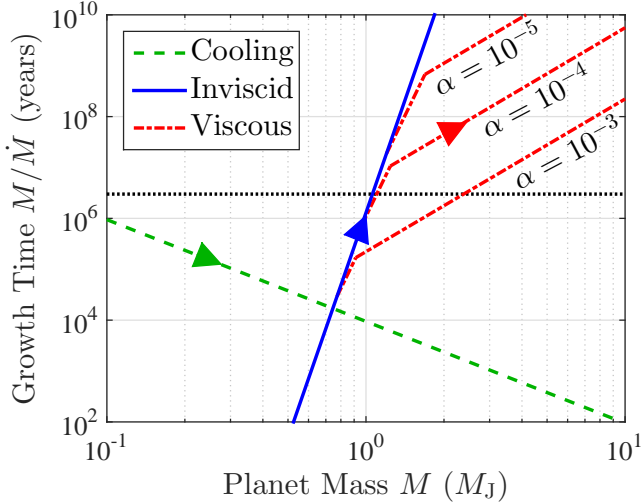}
	\caption{How a gas giant finalizes its mass, from runaway cooling
	to post-runaway accretion in a deep gap. Shown are
	growth (doubling) times 
	of a
	nascent
	gas
	giant
	on a 10 AU orbit (disc temperature $T_0=100\textrm{ K}$)
	in a
	minimum-mass disc
	surrounding a solar mass star. 
	Arrows indicate an example 
	(\citeauthor{ShakuraSunyaev73} viscosity parameter $\alpha=10^{-4}$) evolutionary path.
	The planet first grows at a runaway pace
    set by
	the
	ever-shortening
	Kelvin--Helmholtz cooling time (dashed green line,
	equation \ref{eq:t_cool}). However,
	growth cannot proceed faster than on the Bondi accretion
	time (solid blue and dot-dashed red curves), evaluated using the density
	inside the gap opened by the planet. The Bondi time 
	is initially given by the ``inviscid'' phase during which
	the gap clears without viscous backflow (solid blue
	line, equation \ref{eq:t_m} with $m < m_1$). Later the gap equilibrates
	viscously (in two steps given by the dot-dashed red curve,
	equation \ref{eq:t_m} with $m > m_1$). The planet
	finalizes its mass when the growth time crosses the disc
	lifetime of 3 Myr (dotted horizontal black line).
	For $\alpha \leq 10^{-4}$, the final mass is very nearly
	one Jupiter. For $\alpha = 10^{-3}$, it is between 2--3 Jupiters.}
	\label{fig:growth}
\end{figure}

In the 
next
Section \ref{sec:full} we derive a more accurate expression for $M_{\rm final}$ which takes into account the gap's detailed structure 
($x,w \neq h$) and the disc's finite viscosity. 
Because $w$ increases slowly, we will find that
the gap contrast $\rho_0/\rho$ grows sub-linearly with time.
This leads to a weak power-law (instead of logarithmic) growth for $m(t)$;
see equation \eqref{eq:t_m} and Fig. \ref{fig:growth},
and equation \eqref{eq:final_mass} for the dependence of the final
mass on $t_{\rm disc}$.

Nevertheless, the simplified expression in equation \eqref{eq:simple} exhibits many of the same features of our more carefully derived result in Section \ref{sec:full}.
The final mass depends weakly on $t_{\rm disc}$, has essentially no
dependence on $\alpha$ (by construction in this simplified solution), and depends explicitly on $\rho_0$.
All these features contrast with those of previous
estimates of the final mass. 
For our nominal disc, 
equation \eqref{eq:simple} sets a final mass scale of 
\begin{equation}
M_{\rm final}^{\rm simple}\approx 0.4 M_{\rm J}\left(\frac{a}{10\textrm{ AU}}\right)^{11/14},
\end{equation}
differing only
by a factor of 
$\sim$2
from our more accurate equation \eqref{eq:final_mass}. The scaling of the final mass with distance from the star is also 
similar (11/14 versus 3/4; 
see Section \ref{sec:separation}). One difference
is the scaling with $\rho_0$, which is weaker in the full calculation.

\subsection{Full solution}
\label{sec:full}

\subsubsection{Gap depletion}\label{sec:evolution} 

We focus on low but non-zero viscosities 
for which planet gaps
are so deep that the total torque is no longer dominated
by interaction with gas at a single radial distance $x \sim h$ away
from the planet (as
assumed by \citealt{GoldreichTremaine80} and \citealt{Fung2014}).
Instead, for $\alpha<m^5/h^{14}$,
the torque peaks at two locations: one at $h$, and another
farther away \citepalias{GinzburgSari2018}. Each interaction contributes a factor
of $m^2$ to the density contrast, resulting in a ``two-step''
$m^4$ scaling. This is an asymptotic result that appears
supported by hydrodynamical simulations by \citet[][see in particular the left column of their Fig. 2]{Zhang2018}.
Gaps in discs with higher viscosities obey the classical single-step $m^2$ scaling \citep{DuffellMacFadyen13,Fung2014}. 
We will assess in Section \ref{sec:separation}
the extent to which our assumption that $\alpha < m^5/h^{14}$
is satisfied, after we compute final planet masses.

Gaps evolve toward an equilibrium depth that is
set by a balance between planetary torques which push gas out, and 
viscous torques which ooze material back in.
While \citetalias{GinzburgSari2018} calculate 
this equilibrium and the time required to reach it,
they do not provide gap profiles at earlier times. 
In the Appendix we complete the calculation of \citetalias{GinzburgSari2018} and derive pre-equilibrium, time-varying gap densities. The density
$\rho$ at the bottom of the gap (where the planet resides) relative
to the background density $\rho_0$ is given by
\begin{equation}\label{eq:depletion}
\frac{\rho_0}{\rho}=\begin{cases}
\begin{aligned}
&
m^4h^{-549/49}t^{39/49} & t<t_1\enspace \textrm{(inviscid)}\\
&
m^4h^{-391/35}\alpha^{-4/35}t^{5/7}  & t_1<t<t_2\enspace \textrm{(partial)}\\
&
m^4h^{-53/5}\alpha^{-7/5} & t>t_2\enspace \textrm{(full viscous)}
\end{aligned}
\end{cases}
\end{equation}
where $t$ is the dimensionless time, $t_1=(h^2/\alpha^7)^{1/5}$,  $t_2=(h^4/\alpha^9)^{1/5}$, and it is understood
that the equation applies only for times for which $\rho_0/\rho>1$ 
(this condition turns out to be satisfied for our parameters
because Kelvin--Helmholtz cooling is
long enough for the planet to clear a significant gap; see Fig. \ref{fig:rho}).

Equation \eqref{eq:depletion} is the more accurate version of the simplified equation \eqref{eq:rho_simple}. Three gap-clearing
stages are now delineated. In the first ``inviscid'' stage, the planetary torque 
overwhelms the viscous torque, and so the evolution does not
depend on $\alpha$. This is the regime considered in the previous 
simplified analysis of Section \ref{sec:intuition}; whereas before we
found $\rho_0/\rho \propto m^2/h^5$ following a 
one-step scaling, we now find $\rho_0/\rho \propto m^4/h^{549/49}$
reflecting the two-step nature of low-viscosity,
deep gaps. The scaling with time is now slightly sub-linear because
we have accounted for how the gap width $w$ slowly increases
above $h$. At intermediate times
$t_1 < t < t_2$, the
first density step (closer to the planet) reaches viscous equilibrium
and saturates---hence the dependence on $\alpha$---while the
second step continues to steepen. In the final stage $t > t_2$,
both density steps saturate;
this is the \citetalias{GinzburgSari2018} equilibrium result.

\subsubsection{Growth rates}\label{sec:rates}

\begin{figure}
	\includegraphics[width=\columnwidth]{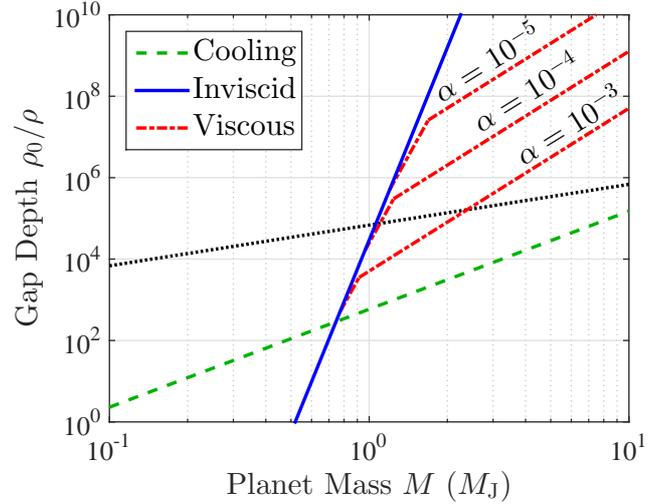}
	\caption{Same as Fig.~\ref{fig:growth}, but showing
	the evolving gap depth, i.e., the 
	contrast between the background nebular density $\rho_0$
	and the density $\rho$ at the bottom of the gap as cleared by a growing gas giant at 10 AU.
    The gap depth during the cooling phase (dashed green line,
    with only the $\alpha=0$ limit plotted for simplicity)
    is calculated by
    inserting equation \eqref{eq:t_cool} into equation \eqref{eq:depletion},
    while the depth during the Bondi-limited phase (solid blue and dot-dashed red curves)
is determined
by inserting 
\eqref{eq:t_m} 
into
\eqref{eq:depletion}. 
The dotted black line, obtained by 
equating $t_{\rm disc} = 3 \textrm{ Myr}$
	to the
	Bondi accretion timescale
	(equation \ref{eq:t_bondi}),
    shows
	how deep the gap must be 
	to stop the growth
	of a planet of given mass $M$ within the lifetime of
	our assumed minimum-mass nebula (Section \ref{sec:disc}).
	Note how this required gap depletion is 
	independent of the specifics of any gap formation theory.
	}
	\label{fig:rho}
\end{figure}

We substitute $\rho_0/\rho(t)$ from equation \eqref{eq:depletion} into equation \eqref{eq:bondi_t0} to obtain planet growth timescales that account for the gap's temporal evolution:
\begin{equation}\label{eq:t_m}
t(m)=\frac{m}{\dot{m}}=\begin{cases}
\begin{aligned}
&
t_0^{49/10}h^{-201/5}m^{147/10}  & m<m_1\enspace \textrm{(inviscid)}\\
&
t_0^{7/2}h^{-143/5}\alpha^{-2/5}m^{21/2}  & m_1<m<m_2 \enspace \textrm{(partial)}\\
&
t_0h^{-38/5}\alpha^{-7/5}m^3  & m>m_2 \enspace \textrm{(full viscous)}
\end{aligned}
\end{cases}
\end{equation}
with $m_1=t_0^{-1/3}h^{58/21}\alpha^{-2/21}$ and $m_2=t_0^{-1/3}h^{14/5}\alpha^{-2/15}$ marking the transitions to
first-step and second-step viscous equilibrium
at times $t_1$ and $t_2$, respectively.
As a reminder, the doubling time $t$ is measured in units of $\Omega^{-1}$, $m\equiv M/M_\star$, $h\equiv H/a$, and $t_0$ is defined in equation \eqref{eq:t0}. 

We plot in Fig. \ref{fig:growth} the planet growth timescales $t(M) = M/\dot{M}$ (with units restored) for different viscosities $\alpha$. 
At the smallest masses (earliest times), accretion is
described by equation \eqref{eq:t_cool}: it is cooling-limited
and proceeds in runaway fashion (dashed green line).
Runaway halts when the planet opens a gap. In the first stage
of Bondi-limited accretion within a gap (equation \ref{eq:t_m},
$m < m_1$, solid blue line), the gap grows inviscidly since 
growth timescales are too short for equilibrium
to be achieved between planetary torques and the initially
weaker viscous torque. 
Departure from the first inviscid stage occurs at different
masses $m_1(\alpha)$. In the subsequent second
($m_1 < m < m_2$, dot-dashed red line, steep segment) and third ($m > m_2$, less steep segment) stages, viscosity
competes more effectively 
and growth timescales are long enough for the gap to reach partial and finally full viscous equilibrium.

The evolution of the gap depth $\rho_0/\rho (M)$ as the planet grows
is shown in Fig.~\ref{fig:rho}, obtained by solving equations
\eqref{eq:bondi_t0} and \eqref{eq:depletion}. 
Even during the earliest cooling-limited growth phase,
a deep gap having $\rho_0/\rho \approx 10$--100 forms.
At low masses ($m<m_1$), the 
gap deepens on the same timescale as
the planet's
mass-doubling time (given first by cooling and then by the inviscid limit);
at high masses ($m>m_2$), the gap 
achieves viscous equilibrium
\citepalias{GinzburgSari2018}. 
By the time the planet is done forming (when the growth time
is as long as $t_{\rm disc}$, i.e., when a given 
blue-red curve intersects with the black line),
the gap depth $\rho_0/\rho \sim 10^5$.
Note how this final depletion factor can be estimated independently
of any gap formation theory: the dotted black line in Fig.~\ref{fig:rho}
is merely given by equating the Bondi-limited growth time
in equation \eqref{eq:t_bondi} to $t_{\rm disc}$. This 
calculation implies that, for our nominal parameters at 10 AU,
a Jupiter-mass planet undergoes its last doubling in mass
in a disc whose gas density is depleted locally
relative to the MMSN by $\sim$$10^5$.

\subsubsection{Final mass}\label{sec:final}

Fig.~\ref{fig:growth} demonstrates that gap clearing
lengthens the Bondi accretion time above
the
Kelvin--Helmholtz 
cooling time
and
ultimately limits a giant planet's growth.
Gap density contrasts increase super-linearly with planet mass
($\rho_0/\rho$ scales as $m^4$ in the low-viscosity limit, and
as $m^2$ at higher viscosities); this dependence
puts an end to runaway accretion by increasing the mass doubling
time with increasing mass.
In the inviscid limit, $t$ scales nearly as $M^{15}$ (equation \ref{eq:t_m};
cf.
the exponential $t(M)$ in our simplified derivation of Section \ref{sec:intuition}).
The inviscid doubling time increases so rapidly that it quickly
exceeds the disc lifetime
$t_{\rm disc}$ (horizontal black line in Fig.~\ref{fig:growth}),
finalizing the planet's mass at
\begin{equation}\label{eq:final_mass}
\frac{M_{\rm final}^{\rm inviscid}}{M_\star}=\left(\Omega t_{\rm disc}\right)^{10/147}\left(\frac{H}{a}\right)^{134/49}\left(\frac{4\uppi G\rho_0}{\Omega^2}\right)^{1/3},
\end{equation}
where 
$\Omega$ is the planet's orbital frequency, $H/a$ is the disc's aspect ratio, and $\rho_0$ is its unperturbed background density.
For our nominal disc parameters at 10 AU,
$M_{\rm final}^{\rm inviscid}\approx 1 M_{\rm J}$. 
Fig.~\ref{fig:growth} shows that this inviscid
final mass applies for $\alpha \lesssim 10^{-4}$, and that 
the final mass increases to $\sim$$2.5 M_{\rm J}$ 
for $\alpha = 10^{-3}$.
That equation \eqref{eq:final_mass} for the final mass
and equation \eqref{eq:mass_thermal} for the thermal mass
scale similarly with $H/a$
implies that our assumption that masses stay sub-thermal
tends to hold regardless of the disc's temperature profile.

Equation \eqref{eq:final_mass} is the more accurate version of equation \eqref{eq:simple}. 
Among the features of the final mass as
given by equation \eqref{eq:final_mass} are that it does not depend on
the (uncertain) disc viscosity and depends only weakly on the disc
lifetime. While the 
simpler version
reproduces these and other qualitative features of the full calculation,
the
quantitative agreement between the two 
is partly coincidental.

\subsubsection{Dependence on orbital distance}
\label{sec:separation}

\begin{figure}
	\includegraphics[width=\columnwidth]{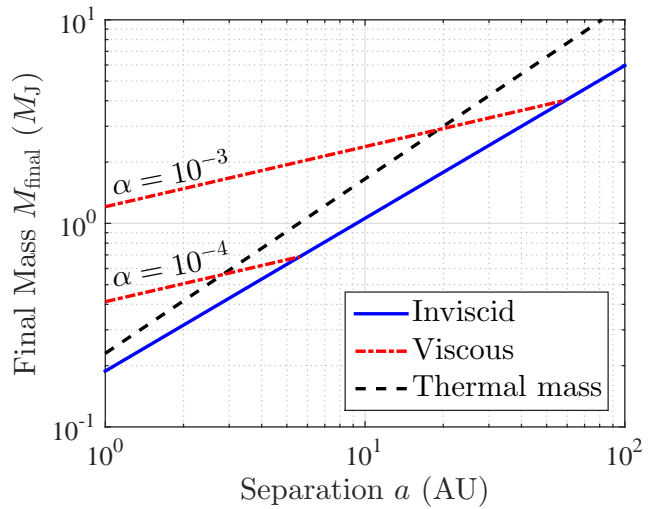}
	\caption{Final planet masses as a function of semi-major axis in 
	our assumed MMSN-like disc (Section \ref{sec:disc}).
	The final mass is calculated by equating the Bondi-limited
	growth time (equation \ref{eq:t_bondi}) to the disc lifetime
	$t_{\rm disc} = 3 \textrm{ Myr}$. The Bondi time depends
	on the density inside the planet's gap. 
	We evaluate this density for two cases.
	In the ``inviscid'' case (solid blue line),
    viscous diffusion times across the gap exceed $t_{\rm disc}$,
    and so the gap clears as if it were inviscid
    (see equation \ref{eq:depletion} for $t < t_1$,
    which leads to equation \ref{eq:final_mass} which is what
    is plotted).
    In the ``viscous'' case (dot-dashed red lines labelled
    by 
    the \citeauthor{ShakuraSunyaev73}
    $\alpha$ parameter), the gap has equilibrated viscously
    within $t_{\rm disc}$ (equation \ref{eq:depletion} for
    $t > t_2$; we have ignored the case $t_1 < t < t_2$ 
    as it applies to a narrow mass range). 
    At large separations $a$, or at low viscosities $\alpha$, 
    the inviscid limit applies.
    The thermal mass (dashed black line) is
    plotted for 
    comparison; our working assumption that
    masses are sub-thermal holds better for larger $a$ and lower $\alpha$.
	}
	\label{fig:distance}
\end{figure}

We present the final planet mass as a function of
distance from the star in Fig. \ref{fig:distance}.
The final mass is computed in one of two ways, either
in the inviscid limit (equation \ref{eq:final_mass})
or when the gap is fully viscously equilibrated (equation
\ref{eq:t_m} for $m>m_2$, with $t$ set to $\Omega t_{\rm disc}$).
We disregard for simplicity the 
intermediate stage of partial equilibrium
since it spans a small mass range and hardly affects
the final mass (see Fig. \ref{fig:growth}).
Fig. \ref{fig:distance} indicates that the inviscid limit applies
at large separations (e.g., $\gtrsim 5$ AU for $\alpha = 10^{-4}$)
and the viscous regime applies at small separations; far from the star,
diffusion times across gaps are so long that the disc behaves
as if it were inviscid.

According to our nominal disc model
(Section \ref{sec:disc}),
$M_{\rm final}^{\rm inviscid}\propto a^{0.75}$.
By comparison,
in viscous equilibrium, 
$M_{\rm final}\propto a^{0.30}$.  
Also plotted for reference in Fig. \ref{fig:distance} 
is the thermal mass $M_{\rm thermal}\propto a^{6/7}\approx a^{0.86}$
(equation \ref{eq:mass_thermal}).
Many of our simplifying approximations (e.g., neglect
of rotation; see Section \ref{sec:caveats})
break down for $M > M_{\rm thermal}$,
and so we see that our final mass estimates are safest
at large distances ($a \gtrsim 3$ AU for $\alpha = 10^{-4}$
and $a \gtrsim 20$ AU for $\alpha = 10^{-3}$),
and then only marginally so.
The ratio of the thermal mass to the inviscid final mass is insensitive to the disc's temperature profile, as both vary similarly with $h$.

Finally, we check whether the low-viscosity condition
$\alpha < m^5/h^{14}$ assumed by our derivation 
(Section \ref{sec:evolution})
is satisfied.
At our nominal 10 AU, $h \approx 0.08$
and $m_{\rm final}^{\rm inviscid} \approx 0.001$,
and so $m^5/h^{14} \approx 2$, validating our assumption
in the case of the inviscid final mass. 
Although pre-final masses are smaller and may in principle violate the 
low-viscosity condition, in practice this is not an issue because
the growth time is an extremely steep function of mass 
(Fig.~\ref{fig:growth}).
Furthermore, for the inviscid final mass, $m^5/h^{14}$
scales only weakly 
with distance, as $a^{-0.25}$, implying that the entire inviscid curve
plotted in Fig.~\ref{fig:distance} is computed self-consistently.
This implies that the viscous curves in Fig.~\ref{fig:distance} 
are also self-consistent, as they yield more massive planets
for which $m^5/h^{14}$ is only larger.
We note that 
$m^5/h^{14}$ is not particularly sensitive to the disc's temperature profile, as the dependence of $m_{\rm final}^{\rm inviscid}$ on $h$ nearly cancels out the
factor of $h^{14}$.

\section{Conclusions and discussion}\label{sec:conclusions}

Gap opening by planets
is a potential
mechanism to stop
runaway gas accretion and thereby
set
final planetary masses.
Contrary to some of the older literature \citep{LinPapaloizou93,Bryden99,IdaLin2004}, this scenario is not restricted
to planets exceeding the thermal mass.
Sub-thermal planets
also
carve out
gaps,
provided
their host discs have low viscosities \citep[e.g.,][]{DuffellMacFadyen13,FungChiang17,GinzburgSari2018}.
With this fact in mind, and motivated further by ALMA observations
that point to low levels of turbulence in discs,
we have
revisited the 
gap starvation hypothesis
and calculated post-runaway masses,
restricting consideration to 
sub-thermal planets
(whose gas envelopes might still be reliably modelled in 1D) 
in low-viscosity discs.
A new feature of our analysis is
an accounting for how
gaps 
may not have enough time to reach
equilibrium depletion levels in 
nearly inviscid environments.
We developed a fully time-dependent theory of gap formation
and coupled it to planetary growth rates.

We identified a purely inviscid regime in which gap depletion
is limited by time and not by viscosity.
By ``inviscid'' we do not mean that the \citet{ShakuraSunyaev73}
viscosity parameter $\alpha$ is 
literally zero.
Rather, gap formation in all discs,
including those with non-zero $\alpha$,
initially proceeds 
{\it as if} gas were inviscid. 
At early times, gravitational torques
from planets dominate viscous torques, and gaps deepen
and widen freely 
with no viscous backflow.
If $\alpha$ is low enough, gaps never evolve beyond this inviscid phase
before the disc dissipates.
We derived in this case an explicit expression for the planet's 
post-runaway mass---the final ``inviscid'' mass---that
does not depend on $\alpha$.
This result is qualitatively different from that of
\cite{TanigawaTanaka16}, who assumed viscous equilibrium.

The inviscid limit applies broadly.
For $\alpha\leq 10^{-4}$
($\alpha\leq 10^{-3}$), 
planets at orbital separations beyond 5 AU (50 AU)
that would otherwise undergo runaway Bondi accretion have their growth halted by gaps that clear inviscidly
(Fig.~\ref{fig:distance}).
For all of the above parameter space,
planet masses remain sub-thermal, consistent with our working assumption, albeit
marginally so.

At 10 AU, growth stops after a few Myr (when the disc expires)
at the final inviscid mass of $1 M_{\rm J}$, as given by equation  
\eqref{eq:final_mass}. The final inviscid mass scales as
$M_{\rm final}^{\rm inviscid} \propto 
t_{\rm disc}^{10/147}\rho_0^{1/3}$, with $t_{\rm disc}$
denoting the disc's lifetime and $\rho_0$ its unperturbed density.
The weak sensitivity to $t_{\rm disc}$ and $\rho_0$ 
and the complete lack of dependence on $\alpha$
imply
that gas accretion onto giant planets is not sensitive
to the details of how gas discs evolve and dissipate.
Note that low viscosities do not necessarily conflict with observed
disc lifetimes,
as modern studies find that 
discs may have laminar (low $\alpha$) midplanes
\citep[e.g.,][]{Pinte2016,Flaherty2017}, with accretion and 
mass loss restricted to surface layers \citep[e.g.,][]{Bai16}.
This justifies our decoupling of $\alpha$ from $t_{\rm disc}$.

Folding together all the factors that depend on
orbital distance $a$ in equation \eqref{eq:final_mass},
we see that the final inviscid mass for giant planets
increases approximately as $a^{0.75}$.
Giant planets are expected to be more massive at larger
distance where disc aspect ratios are larger and
gaps are harder to open (equation \ref{eq:depletion}).
The predicted trend of mass with distance
can be tested observationally, say with
microlensing or direct imaging campaigns.
We emphasize that the prediction
pertains to
giant
planet masses and not to
planet occurrence rates.
Current direct imaging surveys \citep[e.g.][]{Nielsen2019}
indicate that the giant planet
frequency declines outside 10 AU, suggesting
that the controlling factor for giant planet
formation at large distances may not be gas accretion
(the subject of this paper) but instead
the agglomeration of rocky cores massive enough
to undergo runaway \citep[see, e.g.,][and references therein]{Lin2018}. 
Any observational test must also screen out brown dwarfs,
which occur more frequently beyond 10 AU, and which
exhibit other demographic differences with giant planets,
presumably because the former form by gravitational instability
and the latter by core accretion \citep{Nielsen2019}.

We can try, e.g., to compare our predictions against observations of the HR
8799 system. 
The four known planets are spaced between 14
and 68 AU and appear to have roughly equal (model-dependent) masses of
$5-7 M_{\rm J}$ \citep{Bowler2016}. This constant
mass seems to fit the shallower viscous curve ($M_{\rm final}\propto
a^{0.30}$; see Fig. \ref{fig:distance}) better than the inviscid
limit. For higher viscosities lying outside the inviscid limit, equation \eqref{eq:t_m} and Fig. \ref{fig:growth} yield a final mass of several times $M_{\rm J}$, consistent with previous numerical studies \citep[e.g.,][who studied $\alpha=4\times 10^{-4}$ and $\alpha=4\times 10^{-3}$]{Lissauer2009}. 

\subsection{Unresolved issues}\label{sec:caveats}

Notwithstanding the improved theory of gap opening from
\citetalias{GinzburgSari2018} that we have applied to
the problem of stopping runaway, there remain significant
uncertainties in our understanding of deep and wide gaps
in low-viscosity discs. 
Some of 
the theory's
approximations break down when the width of the gap becomes comparable to the
orbital radius, as it does in our models.
Furthermore,
hydrodynamical instabilities 
that are not accounted for in the 1D theory 
(e.g., the
Rossby wave instability; \citealt{Lovelace1999,Li2000})
might in reality
limit 
gap depletion.
For more discussion of
these and other issues,
see 
Section 5.1 of \citetalias{GinzburgSari2018}.

We have assumed that planets, in the act of accreting,
do not consume all the material in their vicinity
and empty their gaps. That is, we have assumed that 
whatever gas is locally accreted 
on the scale of the Bondi radius $R_{\rm B}$ is replenished
by radial transport from farther away, at fast enough rates 
that the gas density 
at $R_{\rm B}$ is maintained 
over the planet's doubling time.
Viscous diffusion offers one means of replenishment,
and is more effective as the gap transitions out
of its inviscid phase, which it does in 
discs with $\alpha\gtrsim 10^{-4}$ 
at about the same time that the planet 
undergoes its last doubling (Fig.~\ref{fig:growth}).
Under these conditions, the planet is in viscous communication
with more-or-less the entire gap, which
retains enough mass at its periphery
to supply a Jupiter's worth of gas at $\sim$10 AU (by construction in the minimum-mass 
nebula).
Another way to replenish gas is 
via hydrodynamic instabilities, not only multidimensional
ones but also those 
in one dimension such as the Rayleigh instability \citep[][and references therein]{TanigawaIkoma2007,YangMenou2010,GinzburgSari2018}. These
can smear away sharp density contrasts
and thereby fill in voids created by the accreting planet.
In general, the problem of determining final giant planet masses is
tied to the problem of how discs transport mass and angular momentum;
the possibility that planet masses are transport-limited deserves
further consideration \citep[see also][and references therein]{TanigawaTanaka16}.

We emphasize that the above
issues are germane to 
any theory that wishes to explain the final masses of gas giants
by gap opening.
As 
Fig. \ref{fig:rho} indicates,
the gap depletion factors necessary to stop accretion
at Jupiter-like masses are rather large, on the order of
$\rho_0/\rho \sim 10^5$, a result that
follows
solely from the Bondi accretion rate and the assumption of a
background minimum-mass disc,
and not from the specific gap-opening theory we have used.
Although we have shown in this paper that such large depletions are achievable within the theory of \citetalias{GinzburgSari2018},
they motivate further work on hydrodynamic instabilities and
radial transport.

Future investigations can also try to lift our restriction
to sub-thermal masses, which as Fig.~\ref{fig:distance} indicates
has prevented us from probing small orbital distances.
Super-thermal masses pose a variety of challenges,
some of which 
are
discussed in \citetalias{GinzburgSari2018}. 
The perturbations that super-thermal planets induce in the disc 
are too strong
to be described by the
linearized equations for wave
excitation \citep{GoldreichTremaine80} and propagation \citep{GoodmanRafikov2001,GinzburgSari2018}.
Furthermore, as $M$ grows above $M_{\rm thermal}$,
and the flow near the planet becomes increasingly rotational
and less spherically symmetric, the gas accretion rate should
deviate from the classical Bondi formula that we have used.
Rotation seems particularly interesting to consider.
In general, whether or not the planet is super-thermal,
the
angular spin velocity of gas 
accreted by the planet
at
its outer edge
is
of order the orbital Keplerian shear $\Omega$. 
For sub-thermal 
planets, $\Omega$ is less than $\omega_{\rm breakup}$, the planet's break-up angular velocity.
Under these circumstances, the planet's envelope
is not rotationally supported; it can contract and allow
more gas from the nebula to be accreted.
But for super-thermal
planets whose atmospheres extend to $R_{\rm H}$,
$\Omega\sim\omega_{\rm breakup}$;
such atmospheres cannot contract
without losing angular momentum.
If they cannot do that---and see, e.g., \citet{Szulagyi2014}
and \citet{Batygin2018} for recent thinking on this
issue---then 
the thermal mass
naturally emerges as a
limit on the planet's final mass (based on considerations that are, to leading order, orthogonal to the
ideas of gap opening
explored in this paper).
Hydrodynamical simulations along these lines have
been conducted by Fung et al.~(2019, in preparation).

\section*{Acknowledgements}

We thank Jeffrey Fung, Eve Lee, Steve Lubow, Ruth Murray-Clay, Mickey 
Rosenthal, Re'em Sari, and Takayuki Tanigawa for discussions. We also thank the anonymous reviewer for comments which improved the paper. SG is supported by the Heising-Simons Foundation through a 51 Pegasi b 
Fellowship.




\bibliographystyle{mnras}
\input{mass.bbl}



\appendix
\section{Time-dependent gaps}\label{sec:gap_depth}

\citetalias{GinzburgSari2018} calculated the equilibrium gap depth in low-viscosity discs. In this 
Appendix
we generalize their calculation
to
derive the time-dependent depth, before equilibrium is reached. We adopt their dimensionless notation in which $G=M_\star=a=1$. 

\citetalias{GinzburgSari2018} found that the gap is composed of two density ``steps''. The first step is generated by waves that originate at a distance $h$ from the planet \citep[this step corresponds to the classical calculation of][]{Fung2014} and deposit their angular momentum at a distance $w_1>h$. Waves that are generated at $w_1$ can further raise the density profile by depositing angular momentum at $w_2>w_1$ \citepalias[see Fig. 2 of][who also show that the third 
and later steps
are
insignificant]{GinzburgSari2018}. We
denote
the surface densities at $h$, $w_1$, and $w_2$ by $\Sigma$, $\Sigma_1$, and $\Sigma_2$, respectively. We 
change 
the notation slightly with respect to \citetalias{GinzburgSari2018}; here we use $\Sigma$
instead of 
their
$\Sigma_0$, and $m$ instead of $\mu$ to denote the
mass ratio.

We begin by analysing the dynamics of the first step.
The torque that the planet generates by interaction with
gas at $h$ is given by $\Sigma m^2/h^3$ \citep{GoldreichTremaine80}. As long as
this
torque remains unbalanced by
viscosity,
gas is displaced
from $w_1$,
which itself expands outward.
The angular momentum required to displace
an
annulus
of
width $w_1$ over its own width is given by $\Sigma_1 w_1^2$ (we approximate $h<w_1<w_2\ll 1$). 
From these considerations we write
\begin{subequations}\label{eq:first_step}
	\begin{equation}\label{eq:sigma1_ineq}
	\frac{\Sigma m^2}{h^3}t=\Sigma_1w_1^2 \,.
	\end{equation}
We also have
	\begin{equation}\label{eq:w1_ineq}
	w_1=\left(\frac{\Sigma_1}{\Sigma}\frac{h^{11}}{m^2}\right)^{1/5} 
	\end{equation}
\end{subequations}
which
relates the wave generation
location 
($h$) and deposition
location 
($w_1$) 
according to equation (11) of \citetalias{GinzburgSari2018}, 
generalizing
the classical \citet{GoodmanRafikov2001} result to waves travelling across a deep gap. Equation \eqref{eq:w1_ineq}
also shows that gaps grow 
wider as they get deeper (as $\Sigma_1/\Sigma$ increases).
We solve 
equations
\eqref{eq:first_step} for the two unknowns:
\begin{subequations}
\begin{equation}\label{eq:sigma1}
\frac{\Sigma_1}{\Sigma}=\begin{cases}
m^2h^{-37/7}t^{5/7} & t<t_1\\
m^2h^{-5}\alpha^{-1} & t>t_1
\end{cases}
\end{equation}
\begin{equation}\label{eq:w1_sol}
w_1=\begin{cases}
(h^8t)^{1/7} & t<t_1\\
(h^6/\alpha)^{1/5} & t>t_1
\end{cases},
\end{equation}
\end{subequations} 
where $t_1\equiv(h^2/\alpha^7)^{1/5}$ marks the transition to a viscously balanced regime ($\Sigma m^2/h^3=\Sigma_1\nu=\Sigma_1\alpha h^2$).
The saturated depth in equation \eqref{eq:sigma1} corresponds to the \citet{DuffellMacFadyen13} and \citet{Fung2014} scaling. 
The depth of the gap grows sub-linearly with time
because of the increasing width $w_1\propto t^{1/7}$.
As explained in Section \ref{sec:evolution}, we disregard 
the earliest 
times
before 
a
gap has
formed
(when 
$\Sigma_1/\Sigma=1$).

We write equations
analogous 
to \eqref{eq:first_step} for the second step, which is generated by the planet's 
interaction with
gas at $w_1$ \citep[we integrate the differential torque there; see, e.g.,][]{LubowIda2010}:
\begin{subequations}\label{eq:second_step}
\begin{equation}
\frac{\Sigma_1 m^2}{w_1^3}t=\Sigma_2 w_2^2
\end{equation}
\begin{equation}\label{eq:w2_ineq}
w_2=\left(\frac{\Sigma_2}{\Sigma_1}\frac{w_1^8h^3}{m^2}\right)^{1/5},
\end{equation} 
\end{subequations}
with equation \eqref{eq:w2_ineq} derived from equation (11) of \citetalias{GinzburgSari2018}.
We solve 
equations \eqref{eq:second_step} to obtain the depth of the second step:
\begin{equation}\label{eq:sigma2}
\frac{\Sigma_2}{\Sigma_1}=\frac{m^2t^{5/7}}{h^{6/7}w_1^{31/7}}=\begin{cases}
m^2h^{-290/49}t^{4/49} & t<t_1 \\
m^2h^{-216/35}\alpha^{31/35}t^{5/7} & t_1<t<t_2 \\
m^2h^{-28/5}\alpha^{-2/5} & t>t_2 \, ,
\end{cases}
\end{equation}
where we substituted $w_1$ from equation \eqref{eq:w1_sol}. At
$t>t_2\equiv(h^4/\alpha^9)^{1/5}$ the second step reaches
viscous equilibrium.

Finally, we combine equations \eqref{eq:sigma1} and \eqref{eq:sigma2} 
to
derive the overall depth of the gap
\begin{equation} \label{eq:final_app}
\frac{\Sigma_2}{\Sigma}=\frac{\Sigma_2}{\Sigma_1}\frac{\Sigma_1}{\Sigma}=\begin{cases}
m^4h^{-549/49}t^{39/49} & t<t_1 \\
m^4h^{-391/35}\alpha^{-4/35}t^{5/7} & t_1<t<t_2 \\
m^4h^{-53/5}\alpha^{-7/5} & t>t_2 \,.
\end{cases}
\end{equation}
As indicated in Section \ref{sec:evolution}, 
equation \eqref{eq:final_app} applies
only
when
$\alpha<m^5/h^{14}$
\citepalias{GinzburgSari2018}. Since $\Sigma=\rho H$ is the density at the bottom of the gap, and $\Sigma_2=\Sigma_0=\rho_0 H$ is the unperturbed density at its 
periphery,
we have derived $\rho_0/\rho=\Sigma_2/\Sigma$
as it 
appears in equation \eqref{eq:depletion}.


\bsp	
\label{lastpage}
\end{document}